# Modeling, Translation, and Analysis of Different examples using Simulink, Stateflow, SpaceEx, and FlowStar

Yogesh Gajula
*Department of Computer Science*
*University of Dayton*
gajulay1@udayton.edu
**(Corresponding Author)**

Ravi Varma Lingala
*Department of Computer Science*
*University of Dayton*
lingalar3@udayton.edu

***Abstract*—This report details the translation and testing of multiple benchmarks, including the Six Vehicle Platoon, Two Bouncing Ball, Three Tank System, and Four-Dimensional Linear Switching, which represent continuous and hybrid systems. These benchmarks were gathered from past instances involving diverse verification tools such as SpaceEx, Flow*, HyST, MATLAB-Simulink, Stateflow, etc. They cover a range of systems modeled as hybrid automata, providing a comprehensive set for analysis and evaluation. Initially, we created models for all four systems using various suitable tools. Subsequently, these models were converted to the SpaceEx format and then translated into different formats compatible with various verification tools. Adapting our approach to the dynamic characteristics of each system, we performed reachability analysis using the respective verification tools.**

## I. INTRODUCTION

Over the last years, there has been big progress in the area of hybrid systems. This growth shows how important it is to have better tools and methods for checking their complexity. This research focuses on testing and confirming key reachability analysis tools: SpaceEx, Flow*, Stateflow, MATLAB-Simulink, and UPPAAL. Hybrid systems have a mixture of discrete as well as continuous dynamics; they pose specific difficulties in the areas of design, modeling, and verification. To face these difficulties, different tools have been presented by researchers, each one designed for specific parts of the analysis of hybrid systems. This study wants to add more understanding about it through a systematic comparison between how well SpaceEx and Flow* work on various benchmarks and types of systems: rectangular, timed, linear continuous, or non-linear. By finding out what is strong and weak in every tool's performance we aim to provide useful knowledge that can improve the way hybrid system design and analysis are done for many different uses.

This research aims to evaluate the effectiveness and reliability of the aforementioned tools for analyzing the reachability of hybrid systems. To achieve this, several performance metrics will be considered, such as accuracy, efficiency, and scalability. This information can ultimately enhance the efficiency and effectiveness of hybrid system design and analysis across various applications.

## II. HYBRID AUTOMATA

In emphasizing the necessity of assessing their precision and reliability, it becomes crucial to ensure the effective utilization of verification tools like Flow* and SpaceEX. One approach to scrutinizing intricate systems involves employing reachability analysis, a method that establishes the spectrum of states a model can attain from a given starting point. Commonly used verification tools such as Flow* and SpaceEX are utilized to confirm the accuracy of reachability analysis. This report focuses on the concept of hybrid automata: a valuable framework for dynamic system modeling. Hybrid automata extend the capabilities of finite-state machines by incorporating real-valued variables that undergo continuous changes within defined time intervals. The benchmarks discussed in Section III are all articulated using hybrid automata. Subsequently, the following section will introduce four benchmarks that utilize verification tools such as Flow*, SpaceEX, Stateflow, UPPAAL, and MATLAB-Simulink and a transformation–translation tool HyST.

## III. TYPICAL BENCHMARKS

In this section, we showcase four distinct benchmarks that have undergone conversion, comprising a Six-Vehicle Platoon, Two-Bouncing Ball, Three-Tank System, and Four-Dimensional Linear Switching. Subsequently, we conduct a reachability analysis for each benchmark using the appropriate verification tools.

### *A. Six Vehicle Platoon*

The platoon system combines a fleet of six vehicles equipped with hybrid powertrains, wherein they travel in close proximity, following a lead vehicle that governs their speed and direction. This arrangement yields several benefits, including improved fuel efficiency, reduced traffic congestion, and enhanced safety. The hybrid powertrain integrates a conventional internal combustion engine with an electric motor and battery pack. The electric motor assists during acceleration, thereby minimizing fuel consumption and pollution. Additionally, the battery can be recharged through a regenerative braking system, capturing, and utilizing energy that would otherwise go to waste when decelerating or coming to a stop. The synergistic interaction of these technologies leads to heightened fuel efficiency and reduced emissions. However, the widespread implementation of a six-vehicle platoon hybrid system faces challenges such as establishing communication protocols for synchronized platoon operation and addressing legal and regulatory barriers that must be overcome before full-scale adoption can be achieved.

The spacing error, or $e_i$, refers to the difference between the car's distance $d_i$, from its predecessor and a reference distance $d_{ref,i}$.

$$e_i = d_i - d_{ref,i}$$

The reachability analysis of the system is seeking to determine a minimal value of $d_{ref,i}$ that guarantees uninterrupted driving. The dynamics of the platoon include:

$$\dot{x} = Ax + Ba_L$$

The state vector is
$x = [e_1, \dot{e}_1, a_{1,} e_2, \dot{e}_2, a_{2,} e_3, \dot{e}_3, a_{3,} e_4, \dot{e}_4, a_{4,} e_5, \dot{e}_5, a_{5,} e_6, \dot{e}_6, a_6]$
where $i's$ is the vehicle acceleration represented by $a_i$.

$$A_c = \begin{bmatrix}
0 & 1 & 0 & 0 & 0 & 0 & 0 & 0 & 0 & 0 & 0 & 0 & 0 & 0 & 0 & 0 & 0 & 0 \\
0 & 0 & -1 & 0 & 0 & 0 & 0 & 0 & 0 & 0 & 0 & 0 & 0 & 0 & 0 & 0 & 0 & 0 \\
1.505 & 4.668 & -3.7734 & -0.7998 & 0.397 & -0.042 & -0.1741 & +0.3516 & -0.0896 & -0.9097 & 0.477 & -0.125 & -1.0099 & 0.417 & -0.043 & -1.0050 & 0.400 & -0.039 \\
0 & 0 & 0 & 0 & 1 & 0 & 0 & 0 & 0 & 0 & 0 & 0 & 0 & 0 & 0 & 0 & 0 & 0 \\
0 & 0 & 0 & 0 & 0 & -1 & 0 & 0 & 0 & 0 & 0 & 0 & 0 & 0 & 0 & 0 & 0 & 0 \\
0.8316 & 3.564 & -0.0694 & 1.0836 & 3.6799 & -2.9396 & -0.555 & 0.1114 & -0.8996 & 1.2196 & 3.9099 & 3.2106 & -1.3136 & 3.6518 & -3.2906 & -1.3154 & 3.6531 & -3.2889 \\
0 & 0 & 0 & 0 & 0 & 0 & 0 & 1 & 0 & 0 & 0 & 0 & 0 & 0 & 0 & 0 & 0 & 0 \\
0 & 0 & 0 & 0 & 0 & 0 & 0 & 0 & -1 & 0 & 0 & 0 & 0 & 0 & 0 & 0 & 0 & 0 \\
0.6932 & 3.493 & -0.0894 & 0.7972 & 3.1968 & -0.0799 & 1.3126 & 3.099 & -3.8856 & 0.9972 & 3.3897 & -0.0896 & 0.7972 & 3.5968 & -0.8816 & 0.7980 & 3.5950 & -0.8830 \\
0 & 0 & 0 & 0 & 0 & 0 & 0 & 0 & 0 & 0 & 1 & 0 & 0 & 0 & 0 & 0 & 0 & 0 \\
0 & 0 & 0 & 0 & 0 & 0 & 0 & 0 & 0 & 0 & 0 & -1 & 0 & 0 & 0 & 0 & 0 & 0 \\
0.7932 & 3.693 & -0.1004 & 0.7972 & 2.98998 & -0.0999 & 1.2343 & 3.897 & -3.9156 & 1.3968 & 3.992 & -3.7806 & 1.9876 & 2.222 & -3.567 & 1.9869 & 2.230 & -.3555 \\
0 & 0 & 0 & 0 & 0 & 0 & 0 & 0 & 0 & 0 & 0 & 0 & 0 & 1 & 0 & 0 & 0 & 0 \\
0 & 0 & 0 & 0 & 0 & 0 & 0 & 0 & 0 & 0 & 0 & 0 & 0 & 0 & -1 & 0 & 0 & 0 \\
0.8972 & 3.129 & -0.0494 & 0.9971 & 3.5433 & -0.0876 & 1.1116 & 3.987 & -3.8956 & 0.0072 & 3.1168 & -0.7657 & 1.2372 & 3.0968 & -1.1276 & 1.2380 & 3.0955 & -1.1269 \\
0 & 0 & 0 & 0 & 0 & 0 & 0 & 0 & 0 & 0 & 0 & 0 & 0 & 0 & 0 & 0 & 1 & 0 \\
0 & 0 & 0 & 0 & 0 & 0 & 0 & 0 & 0 & 0 & 0 & 0 & 0 & 0 & 0 & 0 & 0 & -1 \\
0.8960 & 3.100 & -0.0485 & 0.9900 & 3.5323 & -0.0855 & 1.1108 & 3.955 & -3.8546 & 0.0060 & 3.1130 & -0.7622 & 1.2222 & 3.0928 & -1.1076 & 1.2290 & 3.0885 & -1.1112
\end{bmatrix}$$

$$B_c = \begin{bmatrix} 0 \\ 1 \\ 0 \\ 0 \\ 0 \\ 0 \\ 0 \\ 0 \\ 0 \\ 0 \\ 0 \\ 0 \\ 0 \\ 0 \\ 0 \\ 0 \end{bmatrix}$$

Fig. 1. Matrix for No Communication Problems

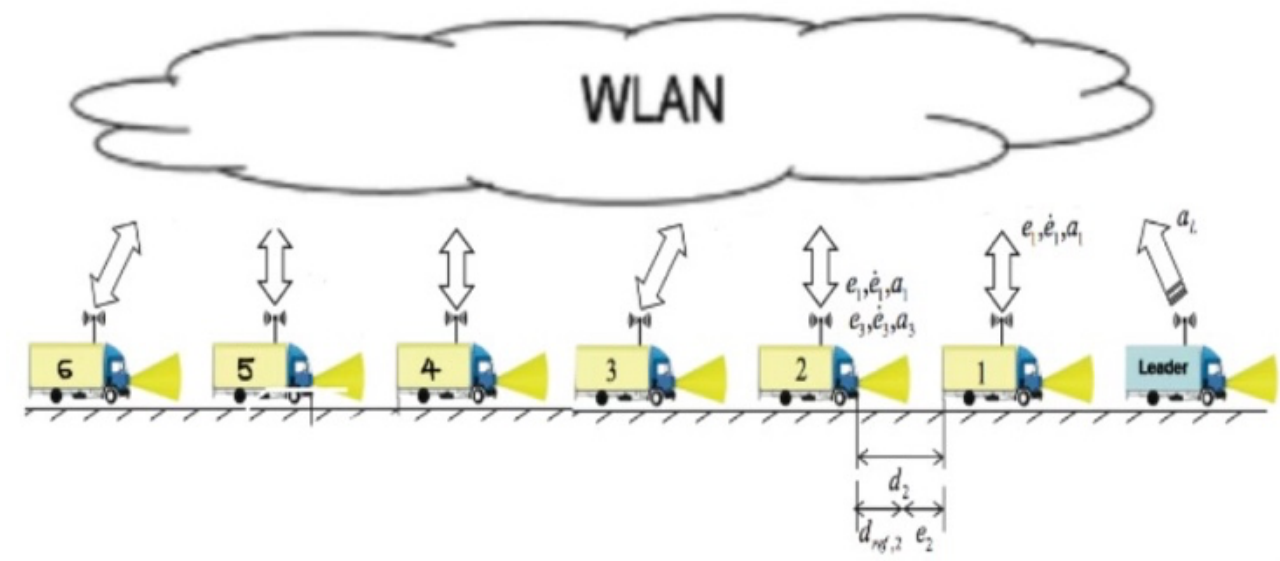

Fig. 2. Visual Representation of Six Vehicle Platoon.

$$A_n = \begin{bmatrix}
0 & 1 & 0 & 0 & 0 & 0 & 0 & 0 & 0 & 0 & 0 & 0 & 0 & 0 & 0 & 0 & 0 & 0 \\
0 & 0 & -1 & 0 & 0 & 0 & 0 & 0 & 0 & 0 & 0 & 0 & 0 & 0 & 0 & 0 & 0 & 0 \\
1.505 & 4.668 & -3.7734 & 0 & 0 & 0 & 0 & 0 & 0 & 0 & 0 & 0 & 0 & 0 & 0 & 0 & 0 & 0 \\
0 & 0 & 0 & 0 & 1 & 0 & 0 & 0 & 0 & 0 & 0 & 0 & 0 & 0 & 0 & 0 & 0 & 0 \\
0 & 0 & 0 & 0 & 0 & -1 & 0 & 0 & 0 & 0 & 0 & 0 & 0 & 0 & 0 & 0 & 0 & 0 \\
0 & 0 & 0 & 1.0836 & 3.6799 & -2.9396 & 0 & 0 & 0 & 0 & 0 & 0 & 0 & 0 & 0 & 0 & 0 & 0 \\
0 & 0 & 0 & 0 & 0 & 0 & 0 & 1 & 0 & 0 & 0 & 0 & 0 & 0 & 0 & 0 & 0 & 0 \\
0 & 0 & 0 & 0 & 0 & 0 & 0 & 0 & -1 & 0 & 0 & 0 & 0 & 0 & 0 & 0 & 0 & 0 \\
0 & 0 & 0 & 0 & 0 & 0 & 1.3126 & 3.099 & -3.8856 & 0 & 0 & 0 & 0 & 0 & 0 & 0 & 0 & 0 \\
0 & 0 & 0 & 0 & 0 & 0 & 0 & 0 & 0 & 0 & 1 & 0 & 0 & 0 & 0 & 0 & 0 & 0 \\
0 & 0 & 0 & 0 & 0 & 0 & 0 & 0 & 0 & 0 & 0 & -1 & 0 & 0 & 0 & 0 & 0 & 0 \\
0 & 0 & 0 & 0 & 0 & 0 & 0 & 0 & 0 & 1.3968 & 3.992 & -3.7806 & 0 & 0 & 0 & 0 & 0 & 0 \\
0 & 0 & 0 & 0 & 0 & 0 & 0 & 0 & 0 & 0 & 0 & 0 & 0 & 1 & 0 & 0 & 0 & 0 \\
0 & 0 & 0 & 0 & 0 & 0 & 0 & 0 & 0 & 0 & 0 & 0 & 0 & 0 & -1 & 0 & 0 & 0 \\
0 & 0 & 0 & 0 & 0 & 0 & 0 & 0 & 0 & 0 & 0 & 0 & 1.2372 & 3.0968 & -1.1276 & 0 & 0 & 0 \\
0 & 0 & 0 & 0 & 0 & 0 & 0 & 0 & 0 & 0 & 0 & 0 & 0 & 0 & 0 & 0 & 1 & 0 \\
0 & 0 & 0 & 0 & 0 & 0 & 0 & 0 & 0 & 0 & 0 & 0 & 0 & 0 & 0 & 0 & 0 & -1 \\
0 & 0 & 0 & 0 & 0 & 0 & 0 & 0 & 0 & 0 & 0 & 0 & 0 & 0 & 0 & 1.2290 & 3.0885 & -1.1112
\end{bmatrix}$$

$$B_n = \begin{bmatrix} 0 \\ 1 \\ 0 \\ 0 \\ 0 \\ 0 \\ 0 \\ 0 \\ 0 \\ 0 \\ 0 \\ 0 \\ 0 \\ 0 \\ 0 \\ 0 \end{bmatrix}$$

Fig. 3. Matrix for Total Disruption of Communication.

*Hybrid Automation:* The platoon brings together various kinds of automation, with continuous changes to its regular conditions. Changes are caused by interruptions in communication. In the most severe situation, which is a dangerous event with no communication among all platoon members

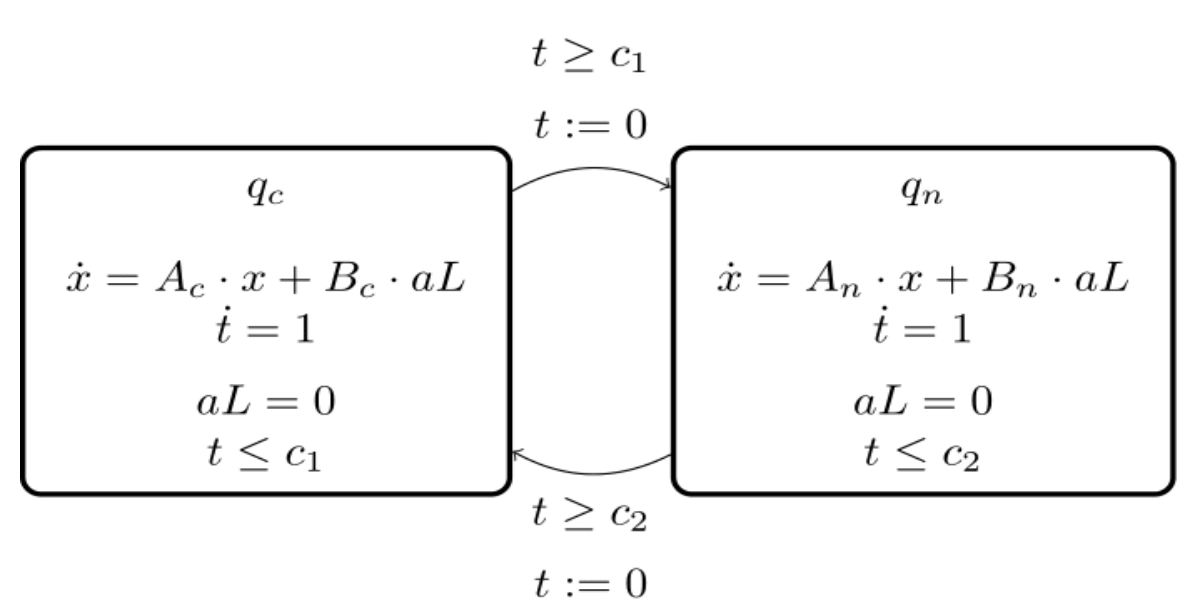

Fig. 4. A Hybrid Automation of Six Vehicle Platoon

Discrete switches, whether spontaneous and true or guard-free transitions, can be modeled to maintain their independence. Alternatively, by adding a clock, the switching process becomes time-dependent.

Otherwise, switching can be made to occur if the algorithm computing the reachable set reaches a fix point. This corresponds to an invariant reachable set in this context. Simulations done during control design assist in figuring out time horizon T, which is when controlled system becomes stable after specific duration of period. For our case study, we have T=12s.

The sets $D_c$ and $D_n$ that match with the state space domains of continuous states $q_c$ and $q_n$ are taken - for this specific application equals to $\mathbb{R}^9$. If not, you can pick a different yet appropriate choice. Also, the user has freedom to decide on values for time constants $c_1$ and $c_2$ of hybrid model.

*Reachability settings:* The initial set is specified by:

- $e_i \in [0.9,1.1]$, $i \in 1,\ldots,6$
- $\dot{e_i} \in [0.9,1.1]$, $i \in 1,\ldots,6$
- $a_i \in [0.9,1.1]$, $i \in 1,\ldots,6$
- $c_1$ , $c_2$ =2 i.e.: jumps taken= 2

The uncertain input for this setup was set to $a_L = 0$ The initial location is $q_c$ and the time horizon is set to T = 12. The set of bad states are the states where e1 ≥ 1.7

*Reachability Analysis*: Using both Flow* and SpaceEx tools reachability analysis for the Six Vehicle Platoon has been conducted. This systematic analysis carefully studied the possible states that the platoon system could reach starting from a specific initial state. By using these tools, we deeply looked at different states that vehicles in platoon can be in. This gave us useful understanding about possible setups and actions. The use of Flow* and SpaceEx together helped in fully studying all reachable states of the system, which improved comprehension regarding dynamics of platoon as well as assessment for its operational parameters.

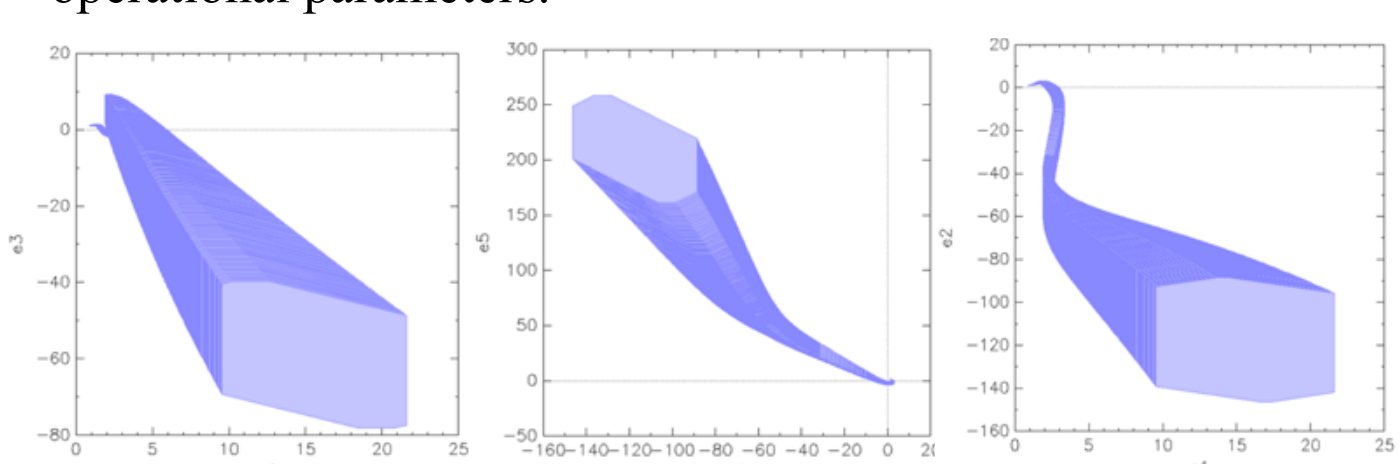

Fig. 5. Reachability analysis of Six Vehicle Platoon using SpaceEx

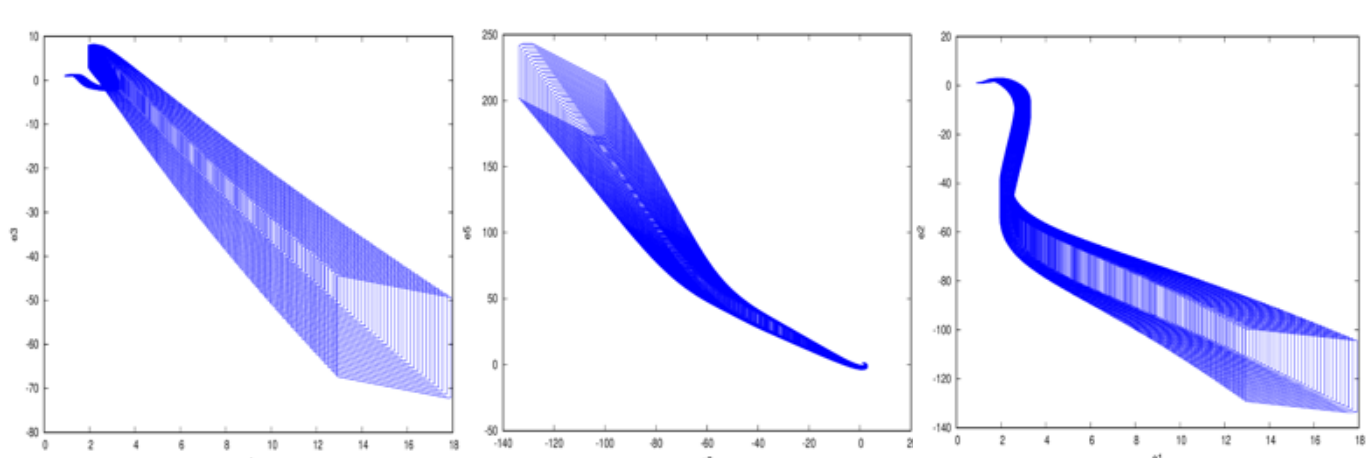

Fig. 6. Reachability analysis of Six Vehicle Platoon using Flow*

*B. Linear Switching System:* This is a piecewise linear system with different controlled continuous dynamics. The matrices A, B, C, and D that describe the modes of the hybrid system are generated randomly using rss() Matlab function. Then these systems created will be stabilized by an LQR controller to ensure convergence towards a stable attracting region. The transitions are determined heuristically using simulations.

The benchmark suggested is made up of 4 modes and 4 transitions. The continuous dynamics in each mode $q_i, i = 1, \ldots, 4$ are displayed by this ordinary differential equation (ODE):

$$\dot{x} = A_i x + B_i u$$

where $x \in \mathbb{R}^4$ is the state vector, u is an input signal that lives in a compact bounded set $U$ and

$$A1= \begin{pmatrix} -0.8036 & 8.7390 & -2.4500 & -8.2700 \\ -8.6218 & -0.5850 & -2.1006 & 3.6000 \\ 2.4510 & 2.2294 & 0.7500 & -3.6922 \\ 1.8299 & 1.9833 & -2.4522 & -1.7316 \end{pmatrix} \qquad B1= \ldots =B4= \begin{pmatrix} -0.0845 \\ 0 \\ 0 \\ 0 \\ -0.7342 \end{pmatrix}$$

$$A2= \begin{pmatrix} -0.8316 & 8.7658 & -2.4744 & -8.2608 \\ -0.8316 & -0.5860 & -2.1006 & 3.6035 \\ 2.4511 & 2.2394 & 0.7538 & -3.6934 \\ 1.5964 & 2.1936 & -2.5872 & -1.6812 \end{pmatrix}$$

$$A3= \begin{pmatrix} -0.9275 & 8.8628 & -2.5428 & -8.2329 \\ -0.8316 & -0.5860 & -2.1006 & 3.6035 \\ 2.4511 & 2.2394 & 0.7538 & -3.6934 \\ 0.7635 & 3.0357 & -3.1814 & -1.4388 \end{pmatrix}$$

$$A4= \begin{pmatrix} -1.4021 & 10.1647 & -3.3937 & -8.5139 \\ -0.8316 & -0.5860 & -2.1006 & 3.6035 \\ 2.4511 & 2.2394 & 0.7538 & -3.6934 \\ -3.3585 & 14.3426 & -10.5703 & -3.8785 \end{pmatrix}$$

Fig. 7. Randomly Generated matrix using rss() in MATLAB

Finally, this Four-dimensional linear switching model has been added to the system for checking its reachability analysis, this has been verified using two different tools, SpaceEx and flow*,to ensure that it behaves as expected and is consistent with the rest of the system.

*Reachability Analysis:* Using both Flow* and SpaceEx tools reachability analysis for the Four-Dimensional Linear Switch has been conducted. This systematic analysis carefully studied the possible states that the platoon system could reach starting from a specific initial state. By using these tools gave us useful understanding about possible setups and actions. The use of Flow* and SpaceEx together helped in fully studying all reachable states of the system, which were similar using both tools. This improved comprehension regarding dynamics of linear switching as well as assessment for its operational parameters.

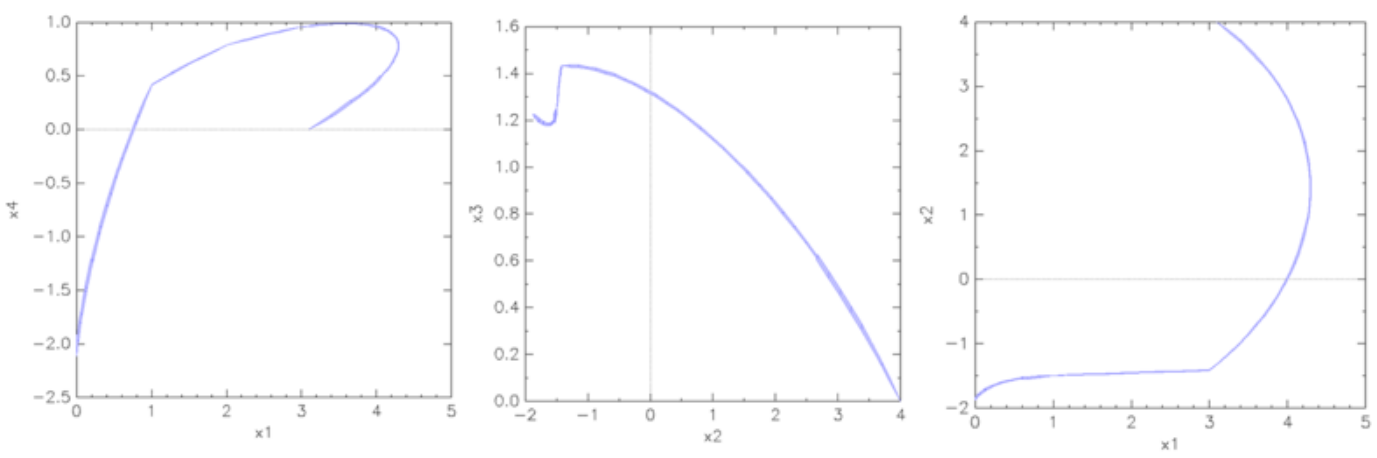

Fig.8. Reachability analysis: 4-Linear Dim Switching using SpaceEx

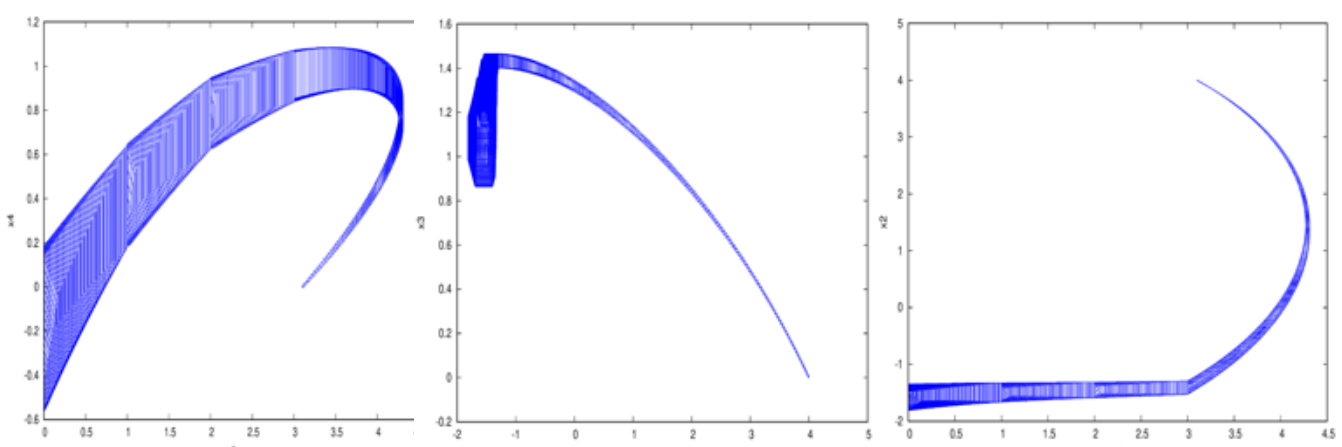

Fig.9. Reachability analysis: 4-Linear Dim Switching using Flow*

*C. Exploring Water Level Control Strategies in Tank Systems:*

The original benchmark model has two tanks. The liquid in the first tank comes from two outside sources: a continuous inflow source and another source with controlled valve valve1, having flows $Q_0$ and $Q_1$ respectively. A drain located at bottom of tank 1 makes the liquid to move into tank 2 with flow $Q_A$ .Tank 2 has two drains. The first drain has a pump to ensure constant liquid outflow $Q_B$ and the second drain's flow $Q_2$ is regulated by an electro-valve valve2. Both valves can be either On or Off. As a result, there exist four feasible discrete modes for the hybrid automaton. The liquid levels in tank i is given by $x_i$.

A new addition, Tank 3, is introduced in this model which receives liquid from Tank 2 through a drain with flow $Q_C$. The flow for the Tank drain is controlled by an electro-valve, valve3, which can also be in the states On/Off.

*Reachability Settings:* This addition of a third tank results in **eight (2^3) discrete modes** for the hybrid automaton. The starting location is *off_off_off* with a time horizon of T = 5s, and a time step of r = 0.1s. The set of bad states are all states where $x_3 = -0.7$.

*Using HyST:* A Source Transformation and Translation Tool for Hybrid Automation Models. The above model is translated from SpaceEX (.xml) format to Flow* (.model). so that reachability can be checked using both tools.

*Reachability Analysis:* For the above-mentioned parameters used tools SpaceEx, HyST, and Flowstar, to recalculate reachable states for this new system which has new configuration. By using these tools gave us useful understanding about possible setups and actions. The use of Flow* and SpaceEx together helped in fully studying all reachable states of the system, which were similar using both tools. This improved comprehension regarding dynamics of Tank systems as well as assessment for its operational parameters.

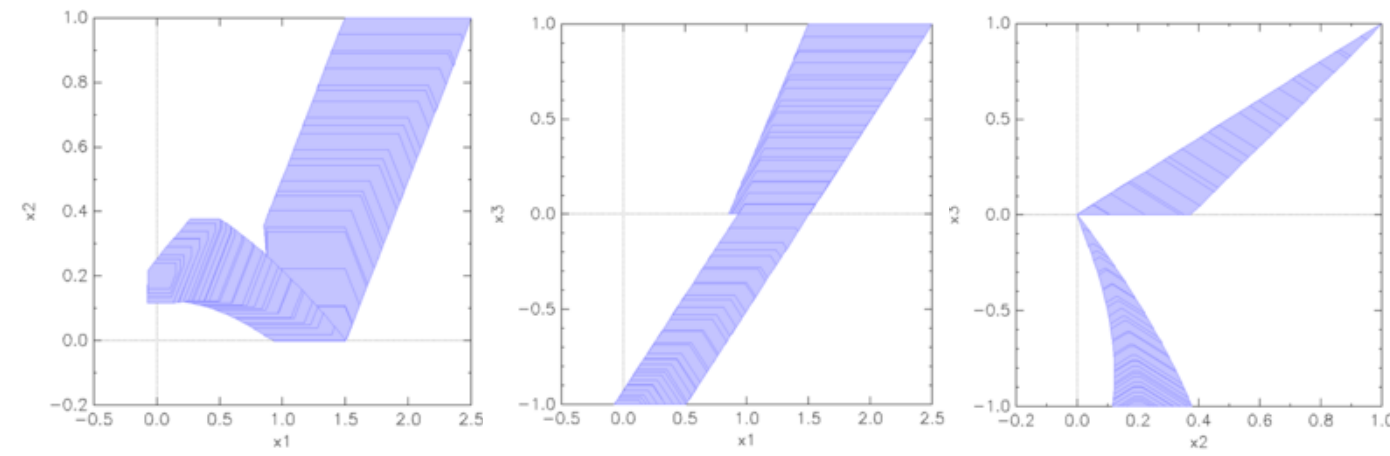

Fig.10. Reachability analysis: Three Tanks System using SpaceEx

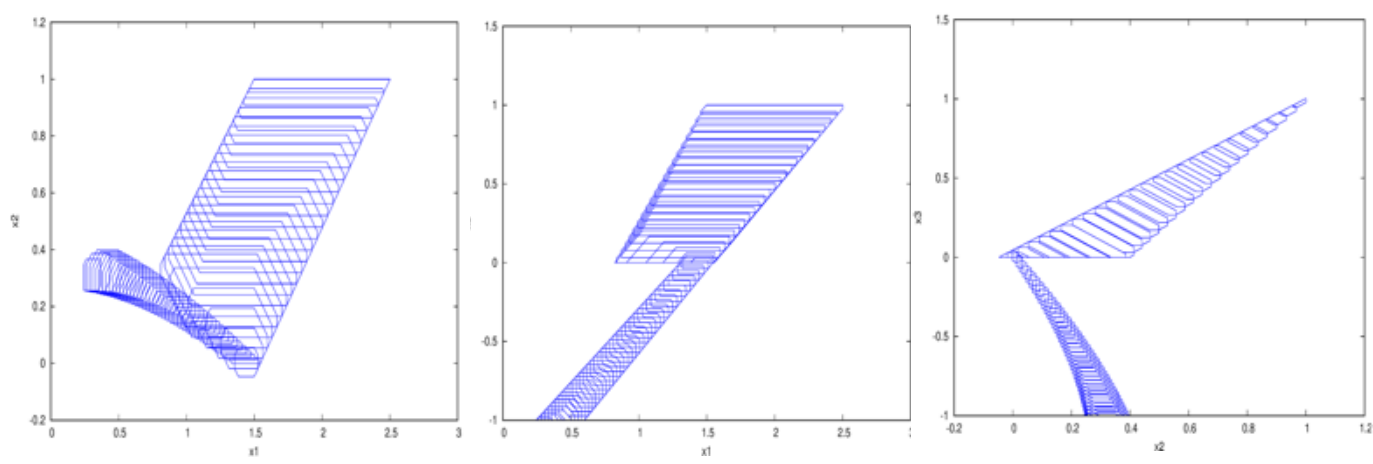

Fig.11. Reachability analysis: Three Tanks System using Flow*

*D. Two Bouncing Ball:*

A well-known example of a bouncing ball is when you drop it from some height and it hits the ground, loses energy but then bounces back up before coming down again. We can represent this physical phenomenon with a hybrid automaton like shown:

The ball's movement in the distinct state is controlled by this differential equation, assuming its mass is m=1 kg: $\dot{x} = v$, $\dot{v} = -9.81$ with x indicating height of ball from ground, v representing its vertical velocity and $9.81ms^{-2}$ being Earth's gravitational force. Ball must bounce every time it hits ground due to invariant $x \geq 0$. Upon reaching the ground, guard $x = 0 \wedge v \leq 0$ of single discrete transition - simulating bounce - ensures bouncing occurs after falling. The energy that is lost due to balls deformation, it is considered by its corresponding reset condition of $v := -c \cdot v$ where $c \in [0, 1]$ signifies constant.

*Reachability Settings:* We consider the initial set:

- $x$, $x1 \in [10; 10.2]$
- $v$, $v1$=0.

We use a time horizon of 40s and set the constant $c = 0.75$. The set of bad states is the set of all states where $v \geq 10.7$.

*Hybrid Automation:* A Hybrid Automaton was used to examine the dynamics of the "Two Bouncing Balls" system. The hybrid system model includes interactions and movements of three bouncing balls in a specified environment. For verification, both Simulink, SpaceEx and Stateflow were employed as tools for analysis. The goal was to understand the system's behavior, which includes its transitions and states that can be reached. This analysis took into consideration the complex dynamics of bouncing balls as well as how they interact with each other in their environment. These tools were used for a deep investigation into how this system behaves and know state transitions, which contributes to overall analysis and verification process.

*Integration Methods:* Two integration methods were utilized to simulate the bouncing balls and observe Zeno behavior. The basic integration method, the Integrator calculates the next state of the system based on its current state and the rate of change. However, it may struggle to accurately capture Zeno behavior due to its limited precision in handling rapid changes. And another integrator method, The Second Order Integrator method considers not only the system's current state and rate of change but also its second derivative, offering higher accuracy and stability, crucial for capturing Zeno behavior effectively.

*Two Integrator's Images:*

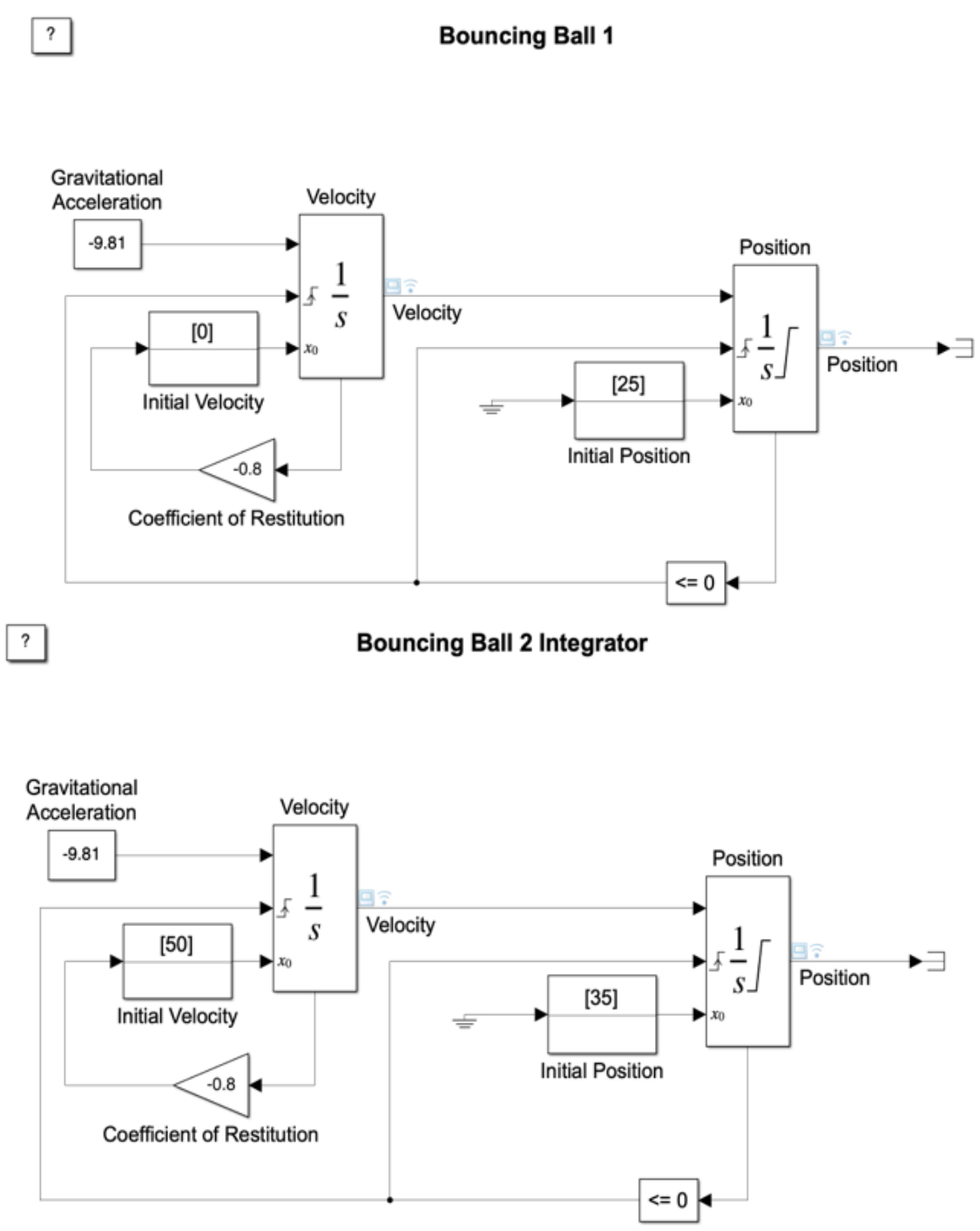

Fig.12. Integrators Images for Ball 1 and Ball 2

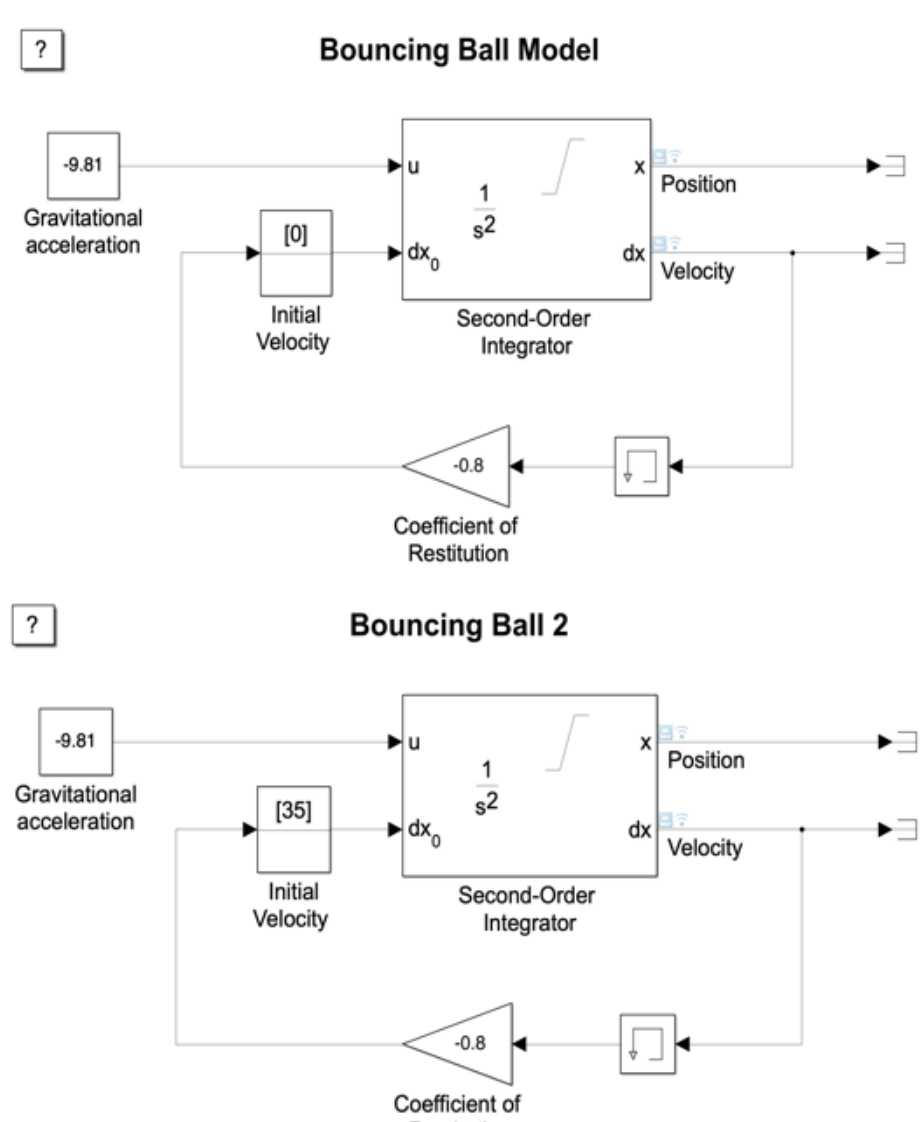

Fig.13. Second Order Integrator Images for Ball 1 and Ball 2

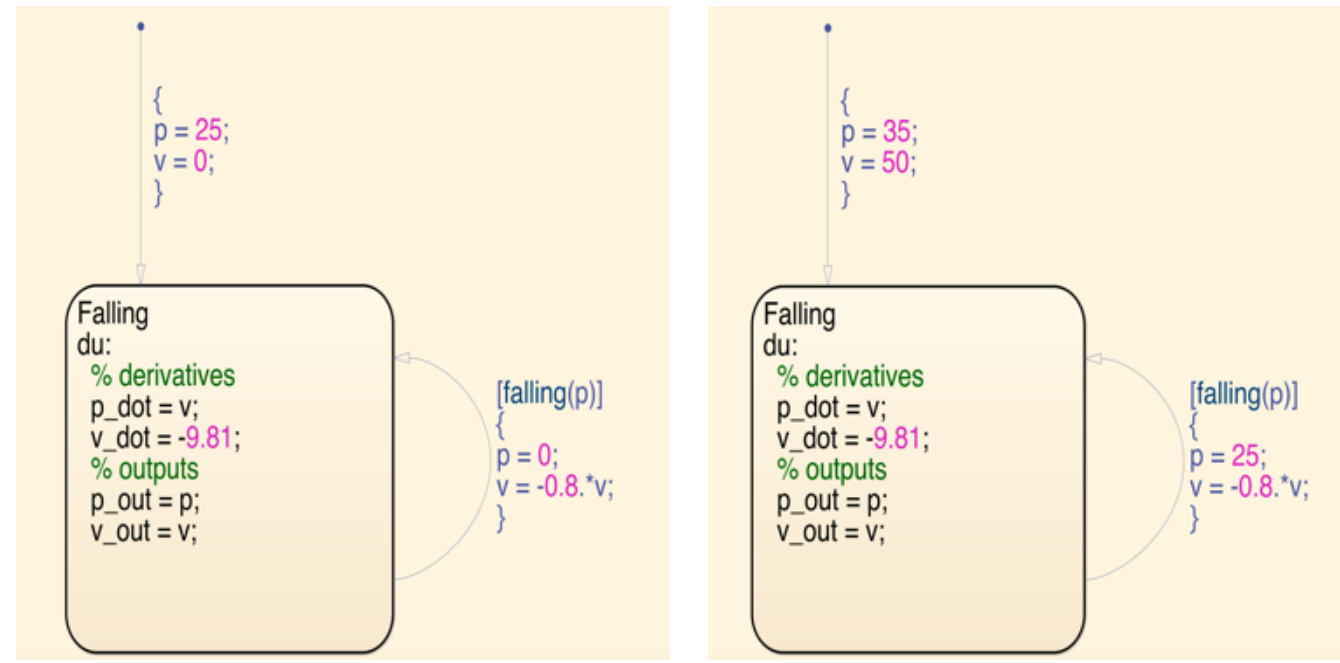

Fig.14. State-flow Charts for Bouncing Ball 1 and Ball 2

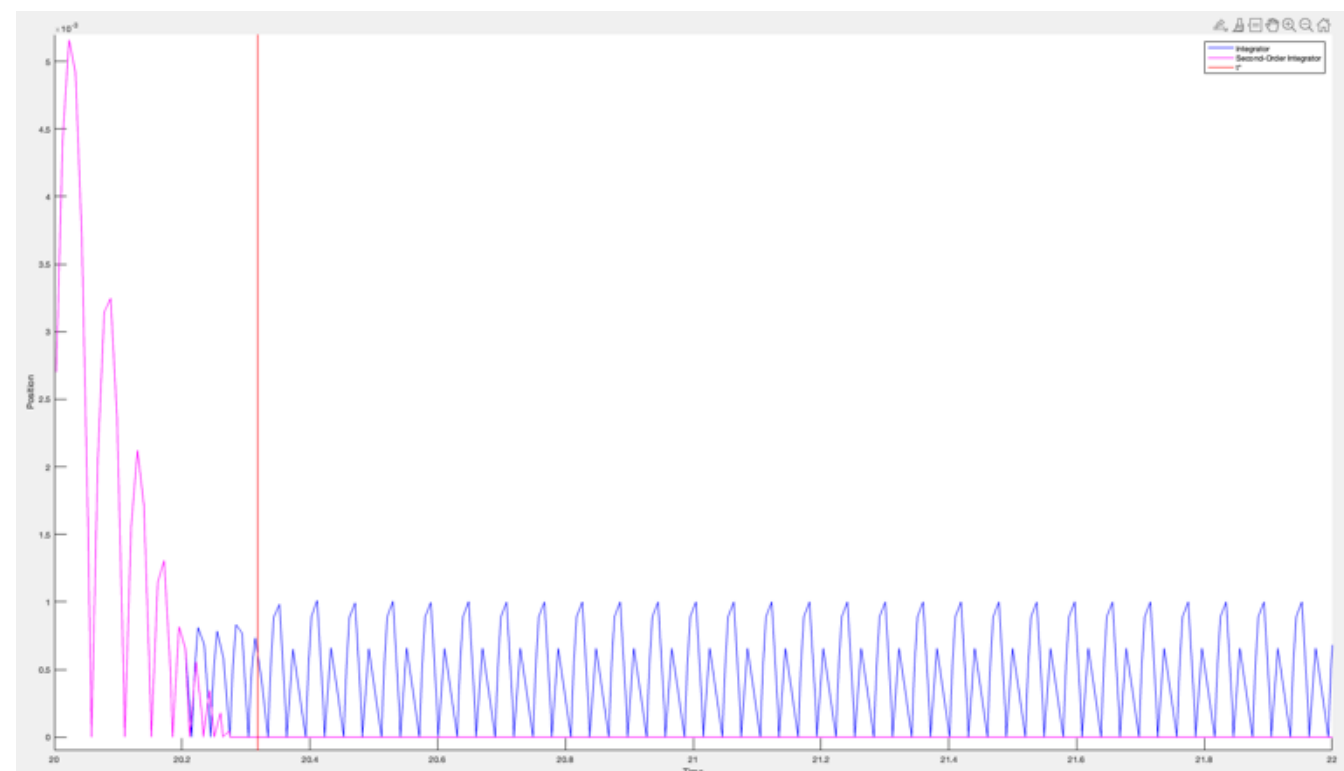

Fig.15. Reachability Analysis for Bouncing Ball 1 using Simulink

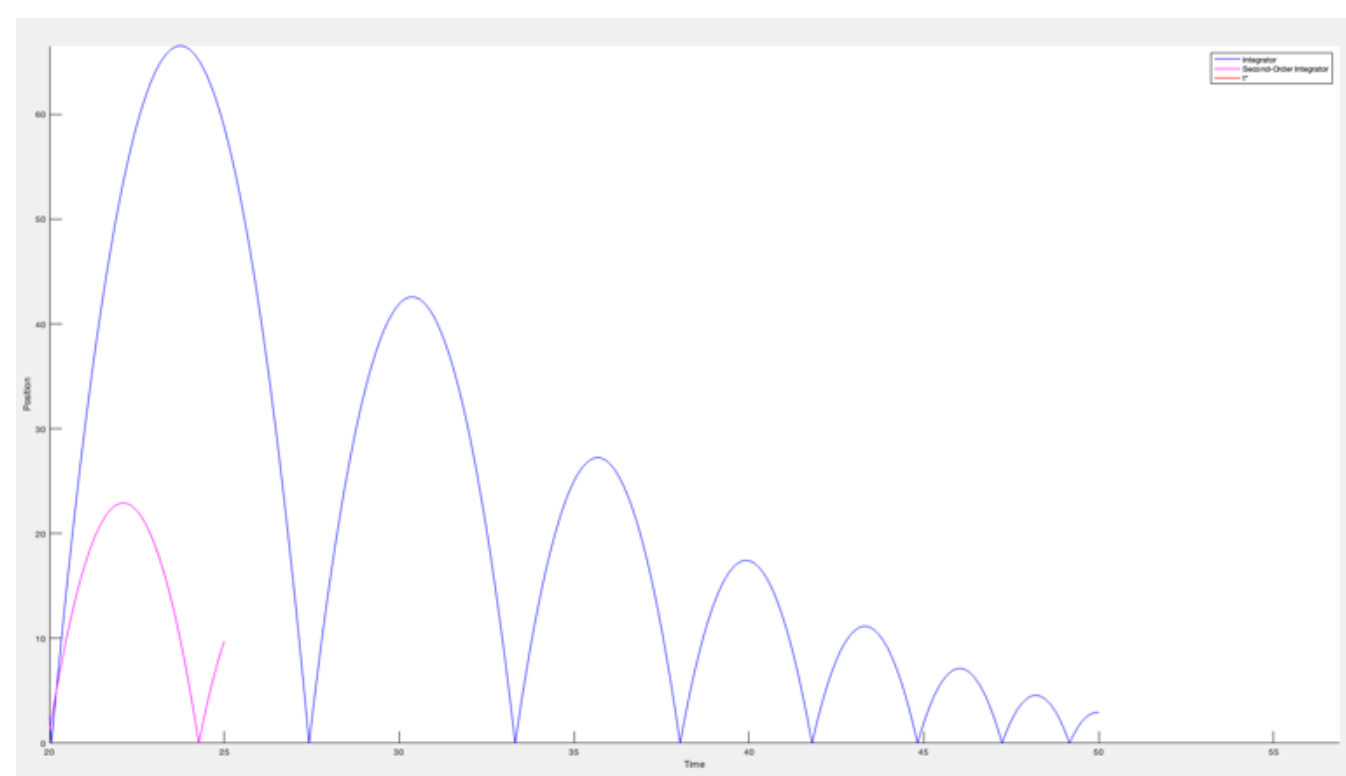

Fig.15. Reachability Analysis for Bouncing Ball 1 using Simulink

*Integrator Observations*: With the standard Integrator, Zeno behavior was observed but with limited clarity. The method struggled to keep pace with the rapid changes, resulting in less precise observations, as shown in the graph above.

*Second Order Integrator Observations:* The Second Order Integrator provided clearer and more detailed observations of Zeno behavior. Its ability to handle rapid changes and subtle variations allowed for a more accurate representation of the system dynamics, as depicted in the graph above.

*Efficiency Analysis:* The efficiency of each integration method in capturing Zeno behavior was assessed based on several criteria.

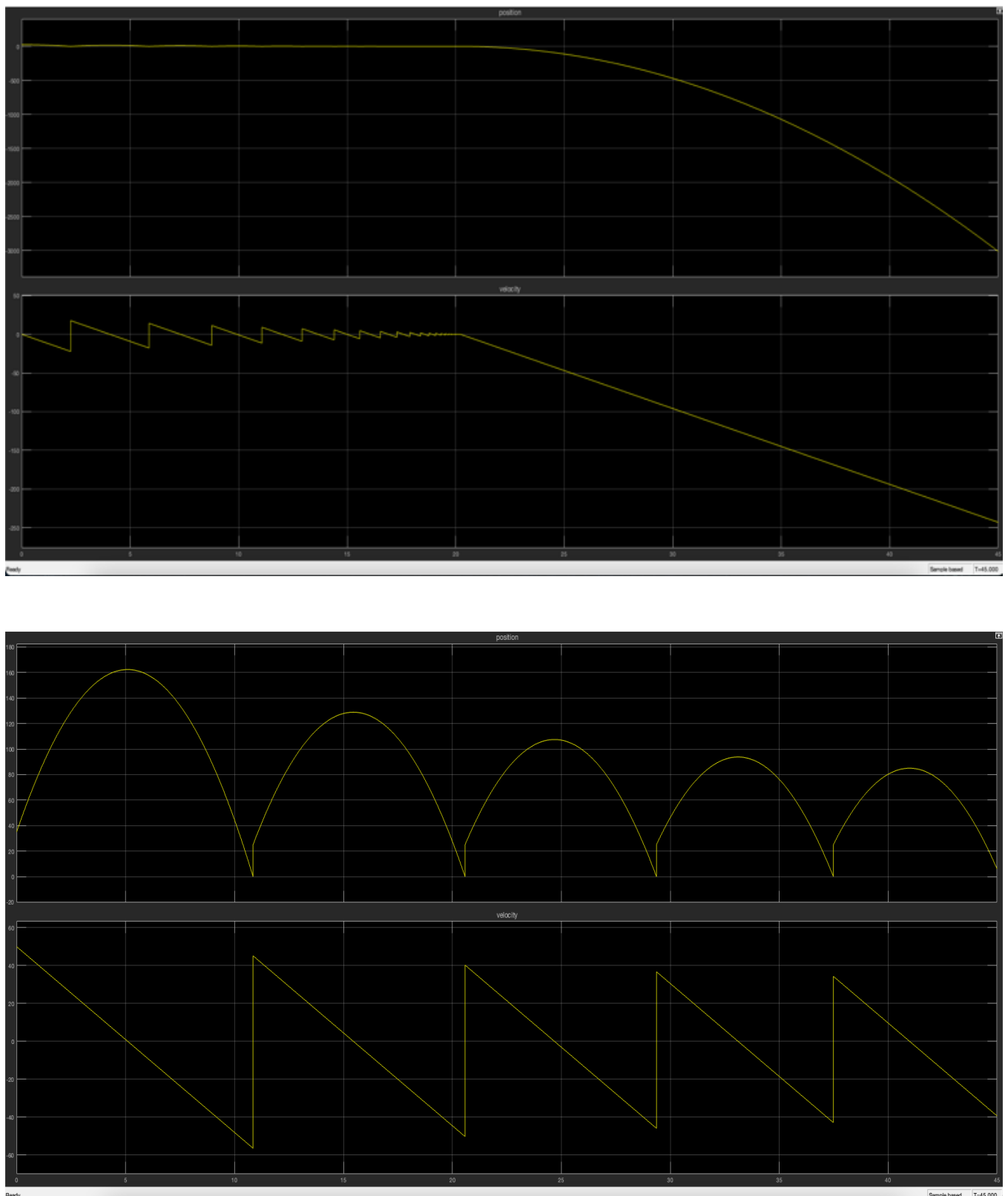

Fig.17. Simulation trajectories when the model runs another time for Bouncing Ball 1 & 2 using Stateflow

In SpaceEX, we implemented the bouncing ball as a model. It showed us that when we drop it from one height and it bounces back to the same height, this is its reachable state. We also learned that while the ball's height decreases and velocity increases, both of them become zero when hitting ground but then bouncing up again; during this time period where velocity decreases and height increases - this process goes on continuously

In conclusion, the Second Order Integrator is recommended for studying Zeno behavior in bouncing balls. Its enhanced accuracy, clarity, and efficiency make it well-suited for simulations involving dynamic systems with rapid, continuous changes. Utilizing this integration method ensures a more precise understanding of Zeno behavior and its implications on system dynamics.

*Reachability Analysis:* For the above-mentioned parameters used tools SpaceEx, MATLAB, Simulink, Stateflow to recalculate reachable states for this new system which has new configuration. By using these tools gave us useful understanding about possible setups and actions. The use of above tools together helped in fully studying all reachable states of the system. This improved comprehension regarding dynamics of Tank systems as well as assessment for its operational parameters. Additionally, the Second Order Integrator offers a higher level of stability, allowing for more reliable predictions of Zeno behavior in bouncing balls.

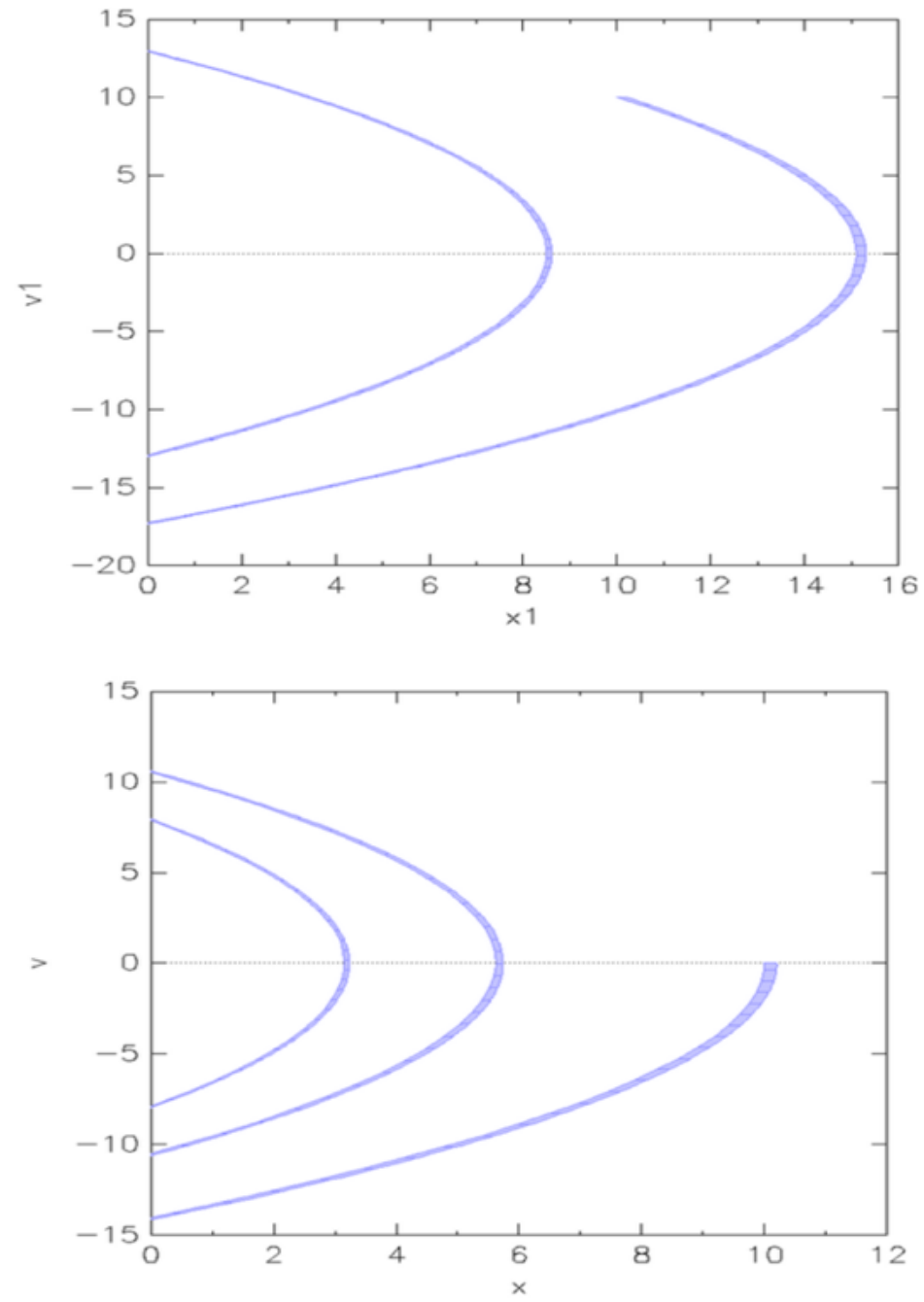

Fig.18. SpaceEx Reachable States for Two Bouncing Balls

## IV. CONCLUSION

In summary, our study highlights the importance of formal verification in confirming complex systems are precise and dependable. We have shown how different tools like Simulink, StateFlow, SpaceEx, Flow*, HyST can be used to validate systems. Changes were made by adding more vehicles in platooning system; extra ball in Bouncing ball model; design and verification process using SpaceEx, Simulink and StateFlow; as well as an additional tank in two tank system which we then analysed with the help of SpaceEx and Flow*. Finally, we made the verification of these systems official and checked if they are reachable. Our results highlight how important formal verification is to guarantee precise and dependable complex system design and analysis.

## FUNDING DECLARATION:

This study was conducted without any funding support.